\documentclass[aps,prb,twocolumn,superscriptaddress,showpacs,floatfix]{revtex4-1}
\usepackage{amsmath,amssymb,amsfonts,float,graphics,epsfig,epstopdf,color,verbatim,tabularx,bm,multirow}
\usepackage{hyperref}
\usepackage{bbold}
\hypersetup{
colorlinks = true,
linkcolor = [rgb]{0.70,0.13,0.13},
citecolor = [rgb]{0.13,0.55,0.13},
urlcolor = [rgb]{0.25,0.41,0.88}}

\DeclareSymbolFont{sfletters}{OML}{cmbrm}{m}{it}
\DeclareMathSymbol{\sfeps}{\mathord}{sfletters}{"22}

\graphicspath{{Figures/}}

\begin{document}

\title{Fermi arcs and pseudogap in a lattice model of a doped orthogonal metal}

\author{Chuang Chen}
\affiliation{Beijing National Laboratory for Condensed Matter Physics and Institute of Physics, Chinese Academy of Sciences, Beijing 100190, China}
\author{Tian Yuan}
\affiliation{State Key Laboratory of Surface Physics, Fudan University, Shanghai 200433, China}
\affiliation{Center for Field Theory
and Particle Physics, Department of Physics, Fudan University,
Shanghai 200433, China}
\author{Yang Qi}
\email{qiyang@fudan.edu.cn}
\affiliation{State Key Laboratory of Surface Physics, Fudan University, Shanghai 200433, China}
\affiliation{Center for Field Theory
and Particle Physics, Department of Physics, Fudan University,
Shanghai 200433, China}
\affiliation{Collaborative Innovation Center of Advanced
Microstructures, Nanjing 210093, China}
\author{Zi Yang Meng}
\email{zymeng@hku.hk}
\affiliation{Department of Physics and HKU-UCAS Joint Institute of Theoretical and Computational Physics, The University of Hong Kong, Pokfulam Road, Hong Kong SAR, China}

\date{\today}

\begin{abstract}
Since the discovery of the pseudogap and Fermi arc states in underdoped cuprates, the understanding of such non-Fermi-liquid states and the associated violation of Luttinger's theorem have been the central theme in correlated electron systems. However, still lacking is a well-accepted theoretical framework to unambiguously explain these metallic states that are clearly beyond Landau's Fermi liquid and Luttinger's theorem of a Fermi surface and electron filling. Here, we design a lattice model of orthogonal metals with fermion and Ising matter fields coupled to topological order and, by solving the model via unbiased quantum Monte Carlo simulation at generic electron fillings, find that the system gives birth to phenomena of the Fermi arc and pseudogap in the single-particle spectrum that go beyond the Luttinger sum rule with broken Fermi surface but no symmetry breaking. The pseudogap and Fermi arcs coexist with a background of a deconfined $Z_2$ gauge field, and we further find that the confinement transition of the gauge field triggers a superconductivity instability and that the hopping of the gauge-neutral fermions brings the "large" Fermi surface back from the Fermi arc state. Our unbiased numerical results provide a concrete model realization and theoretical framework for the coupling between gauge field and fermions and, in the process, generate the rich phenomena of the pseudogap, the Fermi arc, and superconductivity in generic correlated electron systems.
\end{abstract}

\maketitle

\section{Introduction}
 The time-honored Landau's Fermi liquid (FL) theory states that at zero temperature, a Fermi liquid has a closed Fermi surface (FS) marked by the momenta of gapless quasiparticle excitations. 
When the electron number is held fixed, the volume inside the FS is invariant upon interaction; this is the so-called  {\it Luttinger's theorem} (LT)~\cite{Luttinger1960}, and the perturbative argument has been modernized from the topological perspective~\cite{Oshikawa2000,Senthil2004,Paramekanti2004,Heath_2020}. Under these guidelines, the volume inside the FS is conserved even in an interacting FL, and the reduction of the FS must come from the breaking of symmetries.

Given the stringent requirement of LT, the ample experimental observations of correlated electron systems that obviously violate the relation between the volume of a quasiparticle FS and the electron filling therefore pose a serious challenge and show how little we know about interacting metallic states. These systems 
include the Cu-, Fe-, Cr- and Mn-based superconductors~\cite{Loehneysen2007,Keimer2015,YHGu2017,Wu2014,Cheng2017,JGCheng2018}, heavy fermion compounds~\cite{Stewart2001,Custers2003,QMSi2010,Steppke2013,HengcanZhao2019,HQYuanFMQCP2020} and the recently discovered twisted graphene heterostructures~\cite{cao2018correlated,cao2018unconventional,YuanCao2019,ChengShen2019}.  In particular, the experimental observation of Fermi arcs and pseudogap states in underdoped cuprates~\cite{ HongDing1996,Loeser1996,Marshall1996,Norman1998,Kanigel2006,Kondo2013,Hashimoto2014}, where the FS does not form a continuous contour in momentum space but breaks up into disconnected segments and shrinks with decreasing temperature 
to the point nodes below $T_c$, offers the clearest violation of Luttinger's theorem and still awaits a well-accepted explanation.

Many theoretical proposals have been put forward to address the pseudogap and Fermi arcs, such as massless Dirac fermions coupled to gauge fields~\cite{Wen1996,DHKim1997,PLee2006}, fluctuations of the $d$-wave pairing~\cite{Kaminski2015}, competing order with superconductivity~\cite{Norman2005}, finite-temperature lifetime effects~\cite{Yuki}, the fractionalized FL$^{*}$ phase~\cite{Senthil2003,Paramekanti2004,Punk2015,Feldmeier2018,PhysRevB.102.155124} and the Sachdev-Ye-Kitaev (SYK) type of non-Fermi liquid (nFL)~\cite{MaldacenaStanford2016, Sachdev2015,YXWang2020,Hofmann2018, PhysRevResearch.3.013250}. 
The violation of the Luttinger counting has been seen in dynamical cluster approximation (DCA) and high temperature quantum Monte Carlo (QMC) simulations of doped Hubbard model~\cite{KSChen2012,Osborne2020}. However, there exists no lattice realization of a strongly correlated model at generic fillings which can be unambiguously shown to produce the pseudogap and Fermi arc.
Even the FL$^*$ phase still has a closed Fermi surface, enclosing an area that is different from the prediction of LT by half of the Brillouin zone (BZ), with the other half taken by fractionalized excitations to conserve the momentum~\cite{Paramekanti2004,Senthil2003,Senthil2004}.

This is the knowledge gap we would like to fill. Here, we show that the pseudogap and Fermi arc state at generic filling can be observed in a lattice model of correlated electrons with unbiased QMC simulations. The state of the Fermi arc can indeed happen without any symmetry breaking close to the recent observations of similar simulations at half filling~\cite{Gazit2019,ChuangChen2020} and therefore violate LT. Moreover, we discover that the superconducting and normal metal states live in the parameter space neighboring the Fermi arc phase, therefore providing a concrete model realization and theoretical framework for the coupling between the gauge field and fermions and; in the process, we generate the rich phenomena of the pseudogap, the Fermi arc, and superconductivity in generic correlated electron systems.

The lattice model we constructed is composed of fermions and Ising matter fields which are minimally coupled to a $Z_2$ gauge field. Similar models have been proposed as candidates for low-energy effective theories of underdoped cuprates~\cite{Wen1996,DHKim1997,PLee2006,Punk2015,Feldmeier2018}. We find that the Fermi arc phase with a broken FS can transition into a ``large'' FS that respects LT, either via the the enhancement of the hopping of the gauge-neutral composite fermions or the confinement of the $Z_2$ gauge field. The confined FL state at small hopping shows strong superconductivity instability. Our simulations reveal that the Fermi arc state also acquires pseudogap features in the single-particle spectrum and the transitions from Fermi arc to large FS look continuous. These findings offer an unbiased realization of the Fermi arc and pseudogap phenomena from a lattice model at generic fillings.

\begin{figure}[htp!]
	\includegraphics[width=0.9\columnwidth]{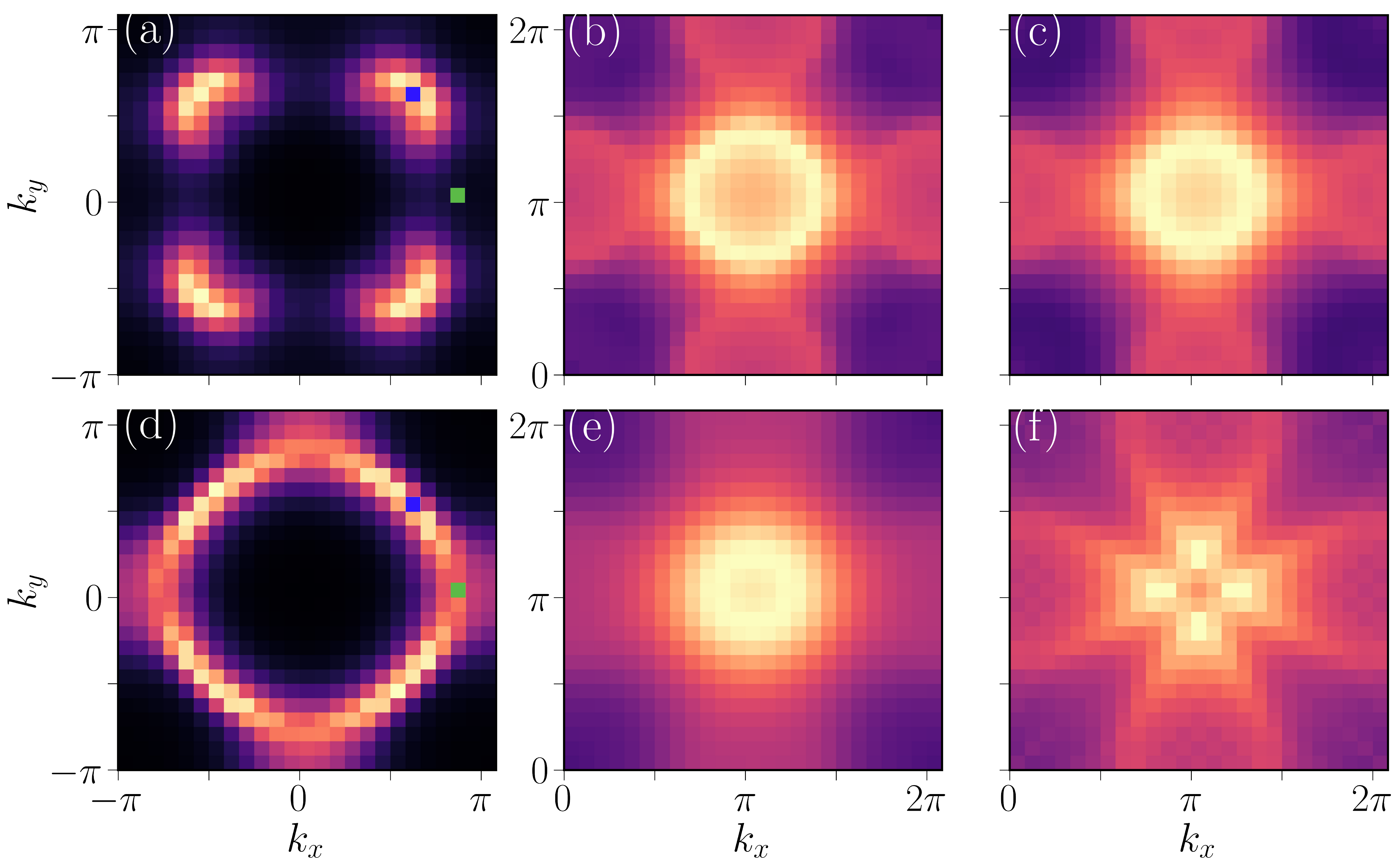}
	\caption{Quasiparticle fraction and spin susceptibility. The parameters are $L=24, T=0.05, g=0.5$ with $t=0.3$ (filling $n=1.12$) in (a) and (b) and $t=1.0$ (filling $n=1.11$)  in (d) and (e). The blue and green squares are the representative momenta along nodal and antinodal directions respectively, where quasiparticle fractions are extrapolated in Fig.~\ref{fig:fig4} (b). The Fermi arcs are shown in (a), and the large FS is shown in (d). (b) shows the $c$-fermion spin susceptibility $\chi(\mathbf{q},\omega=0)$ inside the Fermi arc phase, and (c) shows it in free doped Dirac cones at the same filling as in (b). They acquire the same magnetic response meaning that the Fermi arc phase has a hidden FS of $f$ fermions with the same shape of a doped Dirac cone. (e) shows the $c$-fermion spin susceptibility for the large FS case, and (f) shows that of the free Hamiltonian
		$H = -\sum_{ i,j } (t_{ij} f^{\dagger}_{i,\alpha} f_{j,\alpha} + \text{H.c.}) -\mu\sum_{i}f^{\dagger}_{i,\alpha}f_{i,\alpha}$ with nearest-neighbor (NN) and next-nearest-neighbor (NNN) hoppings.}
	\label{fig:fig1}
\end{figure}

\section{Model}
\begin{figure}[htp!]
	\includegraphics[width=0.9\columnwidth]{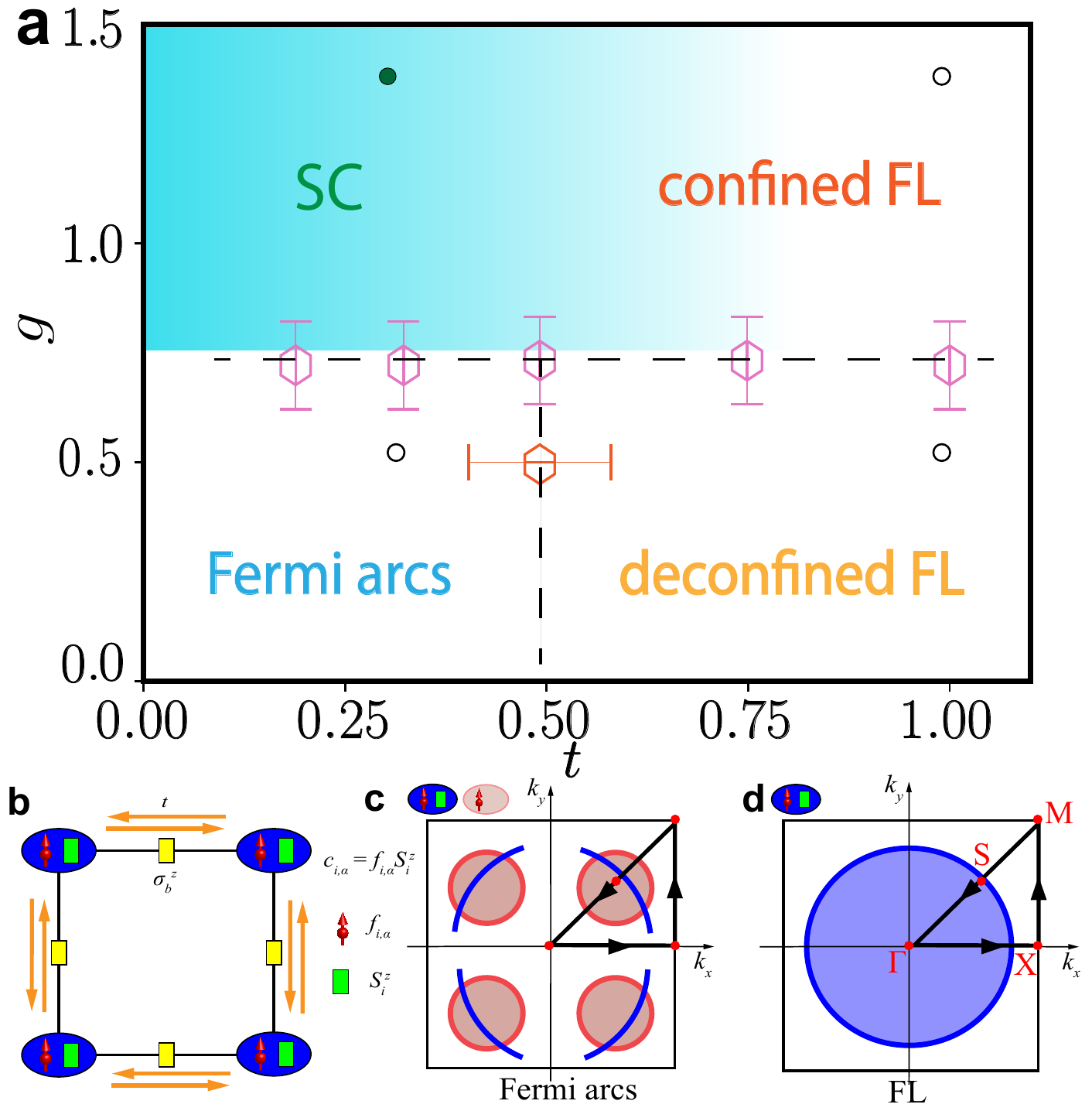}
	\caption{(a) $t-g$ phase diagram. At $g<g_c$ and $t<t_c$ (where $g_c$ and $t_c$ are denoted by the horizontal dashed line and vertical solid line, respectively), the Fermi arc state is obtained by doping the orthogonal semimetal. At $g<g_c, t>t_c$, the deconfined FL phase with $Z_2$ topological order coexists with large FS. At $g>g_c$, the $Z_2$ gauge field is confined and the fermion then forms a conventional confined FL with large FS. It is unstable towards $s$-wave superconductivity (SC) at small $t$. The solid and open circles are the parameters where we study the pairing instability, where the solid circle has a finite $T_c$. (b) The model on a square lattice. There are composite fermions (blue ellipses) $c_{i,\alpha}=f_{i,\alpha}S^{z}_{i}$ on each site $i$, composed of an orthogonal fermion field $f_{i,\alpha}$ and Ising matter field $S^{z}_{i}$. The $Z_2$ gauge field $\sigma^{z}_{b}$ lives on the bond.    (c) The red pockets are the hidden $f$-fermion FS inside the Fermi arc phase $(g<g_c, t<t_c)$, undetectable through single-particle spectra, but they can be inferred from the spin susceptibility data [Fig.~\ref{fig:fig1}(b)]. The blue arcs are the Fermi arc in this phase that can be detected from experiments as shown by the single-particle spectra in Fig.~\ref{fig:fig4}(c). (d) The blue circle is the $c$-fermion large FS inside the confined FL phases, as shown by the single-particle spectra in Fig.~\ref{fig:fig4}(d). High-symmetry points inside the BZ ($\Gamma$, X, M, and S) are denoted.}
	\label{fig:fig2}
\end{figure}

Our model, inspired by the orthogonal fermion constructions~\cite{Nandkishore2012,Reuegg2010} and subsequent lattice model simulations~\cite{Gazit2019,ChuangChen2020}, has the following Hamiltonian on a two-dimensional (2D) square lattice, $H=H_{f}+H_{z}+H_{g}+H_{c}$, where as shown in Fig.~\ref{fig:fig2},
$H_f = -\sum_{\langle i,j \rangle} (f^{\dagger}_{i,\alpha} \sigma^{z}_{b_{\langle i,j \rangle}}f_{j,\alpha} + \text{H.c.}) -\mu\sum_{i}f^{\dagger}_{i,\alpha}f_{i,\alpha}$ describes the orthogonal fermion, with nearest neighbor (NN) hopping amplitude set at unity, the chemical potential $\mu$ and the spin $\alpha=\uparrow,\downarrow$;
$H_{z} = J \sum_{\langle i,j \rangle} S^{z}_{i} \sigma^{z}_{b_{\langle i,j \rangle}} S^{z}_{j} - h \sum_{i} S^{x}_{i}$ is the Ising matter field, with NN antiferromagnetic interaction $J=0.1$ and the transverse field $h=0.25$ to promote quantum fluctuations;
$H_{g} = K \sum_{\square}\prod_{b\in\square} \sigma^{z}_{b} - g\sum_{b} \sigma^{x}_{b}$ describes the $Z_2$ gauge field, with $K=1$ such that $\pi$-flux per plaquette $\square$ is favored, and $g$ triggers the confinement transition of the gauge field; and
$H_{c} = -t \sum_{\langle i,j\rangle} f^{\dagger}_{i,\alpha}S^{z}_{i}f_{j,\alpha}S^{z}_{j} + \text{H.c.}$ defines the NN hopping of the physical -- gauge neutral -- composite fermion $c^{\dagger}_{i,\alpha} (c_{i,\alpha}) = f^{\dagger}_{i,\alpha} S^{z}_{i} (f_{i,\alpha}S^{z}_{i})$ [denoted as blue ellipses in Fig.~\ref{fig:fig2}(b)], and we tune $t$ to enhance the $c$-fermion hopping such that the Fermi arc to 'large' FS transition can be realized.
The QMC implementation of this model is present in detail in Appendix ~\ref{app:QMC}.

As shown in Refs.~\cite{ChuangChen2020,Gazit2019}, at half filling of the $f$ electrons ($\mu=0$), the $\pi$ flux of the $Z_2$ gauge field produces an orthogonal semimetal (OSM) state in which the FS of the $c$ fermions reduces to four Dirac points located at the nodal point $(\pm\frac{\pi}{2},\pm\frac{\pi}{2})$ of the BZ. Here, we start from the orthogonal semimetal state but tune the chemical potential $\mu$ away from half filling.

The most striking results are shown in Fig.~\ref{fig:fig1}. With parameters $L=24, T=0.05, g=0.5$, we first contrast the Fermi arc phase in Fig.~\ref{fig:fig1} (a) ($t=0.3$) with the large FS phase in Fig.~\ref{fig:fig1} (d) ($t=1$). The $c$-fermion spectral function can be approximated via its Green's function as $A(\mathbf{k},\omega=0) \propto \beta G(\mathbf{k},\beta/2)$. The chemical potential in both cases is $\mu=1.2$ and their corresponding fillings are $n=1.12$ and $n=1.11$. At such fillings, the large FS in Fig.~\ref{fig:fig1}(d) respects LT, and the Fermi arc in Fig.~\ref{fig:fig1}(a) certainly violates it. We have also performed the finite size extrapolation of quasiparticle fractions, shown in Appendix~\ref{app:FS}.

Figures.~\ref{fig:fig1}(b) and ~\ref{fig:fig1}(e) show the magnetic response of the Fermi arc and deconfined FL in Figs.~\ref{fig:fig1}(a) and ~\ref{fig:fig1}(d). We measure the gauge-neutral magnetic susceptibility of the $c$ fermions, $\chi(\mathbf{q},\omega=0)=\frac{1}{\beta N}\int^{\beta}_{0}d\tau \sum_{i,j} e^{i\mathbf{q}\cdot \mathbf{r}_{ij}}\langle (n^{\uparrow}_{i,c}-n^{\downarrow}_{i,c})(\tau)(n^{\uparrow}_{j,c}-n^{\downarrow}_{j,c})(0)\rangle$.
It is interesting to see that in both cases the magnetic responses are strongest in the vicinity of $(\pi,\pi)$ (the ring-shaped circles), which means that both cases acquire a similar shape of FS giving rise to a similar magnetic response, only that in the former it is the gauge-dependent, hidden FS of $f$ fermions but in the latter, it is the FS of gauge-neutral $c$ fermions.

To make the contrast clearer, in Figs.~\ref{fig:fig1}(c) and ~\ref{fig:fig1}(f) we plot the magnetic susceptibility for free fermions. In Fig.~\ref{fig:fig1}(c) we compute the $\chi(\mathbf{q},\omega=0)$ for doped Dirac fermions, which is generated by the $\pi$-flux square lattice with $H_f$ only, and replace the $Z_2$ gauge field therein by static phase factor $e^{i\frac{\pi}{4}}$. At the filling $n=1.13$ we observe almost identical $\chi(\mathbf{q},\omega=0)$ to that of Fig.~\ref{fig:fig1}(b); this again implies that the Fermi arc state actually acquires hidden Fermi pockets of $f$ fermions with the same shape of doped Dirac cones and consequently gives rise to the same magnetic response, although the actual FS of doped orthogonal metal is the broken Fermi arcs, which violates LT.

Lastly, in Fig.~\ref{fig:fig1}(f), we plot the magnetic susceptibility of free fermions with a large FS, obtained from $H=-\sum_{i,j}(t_{i,j,\alpha}f^{\dagger}_{i,\alpha}f_{j,\alpha}+\text{H.c.}) - \mu \sum_{i}f^{\dagger}_{i,\alpha}f_{i,
\alpha}$, and we tune the hopping $t_{i,j}$ with $t_{\text{NN}}=1.0$ and $t_{\text{NNN}}=0.1$, and $\mu=-0.5$ such that this free system will also give rise to a FS similar to that of Fig.~\ref{fig:fig1}(d). The $\chi(\mathbf{q},\omega=0)$ of such a large FS is shown with the bright response close to $(\pi,\pi)$.

\section{Phase diagram}
With the Fermi arc and large FS phases seen, we move on to the entire $t-g$ phase diagram, as shown in Fig.~\ref{fig:fig2}(a). It contains three different phases:  the Fermi arc with pseudogap, the deconfined FL, and the confined FL (an overview of the three phases can be found in Appendix~\ref{app:PHASE}). The transition between Fermi arc and deconfined FL phases is triggered by the composite fermion hopping $t$, as shown in Figs.~\ref{fig:fig1}(a) and ~\ref{fig:fig1}(d) with $t=0.3$ and $t=1$, respectively. A heuristic understanding of the FS of these two phases is shown in Figs.~\ref{fig:fig2}(c) and ~\ref{fig:fig2}(d). Inside the Fermi arc phase, the $f$ fermion acquires the FS of doped Dirac cones, denoted by the red pockets, but the FS of the $c$ fermion here is only the broken solid blue arcs in Fig.~\ref{fig:fig2}(c). However, when the $c$-fermion hopping is enhanced, the system enters a metallic phase with large FS, shown as the solid blue circle in Fig.~\ref{fig:fig2}(d). Inside this phase, the $Z_2$ gauge field is still deconfined, coexisting with the large FS.

\begin{figure}[htp!]
\begin{center}
\includegraphics[width=0.9\columnwidth]{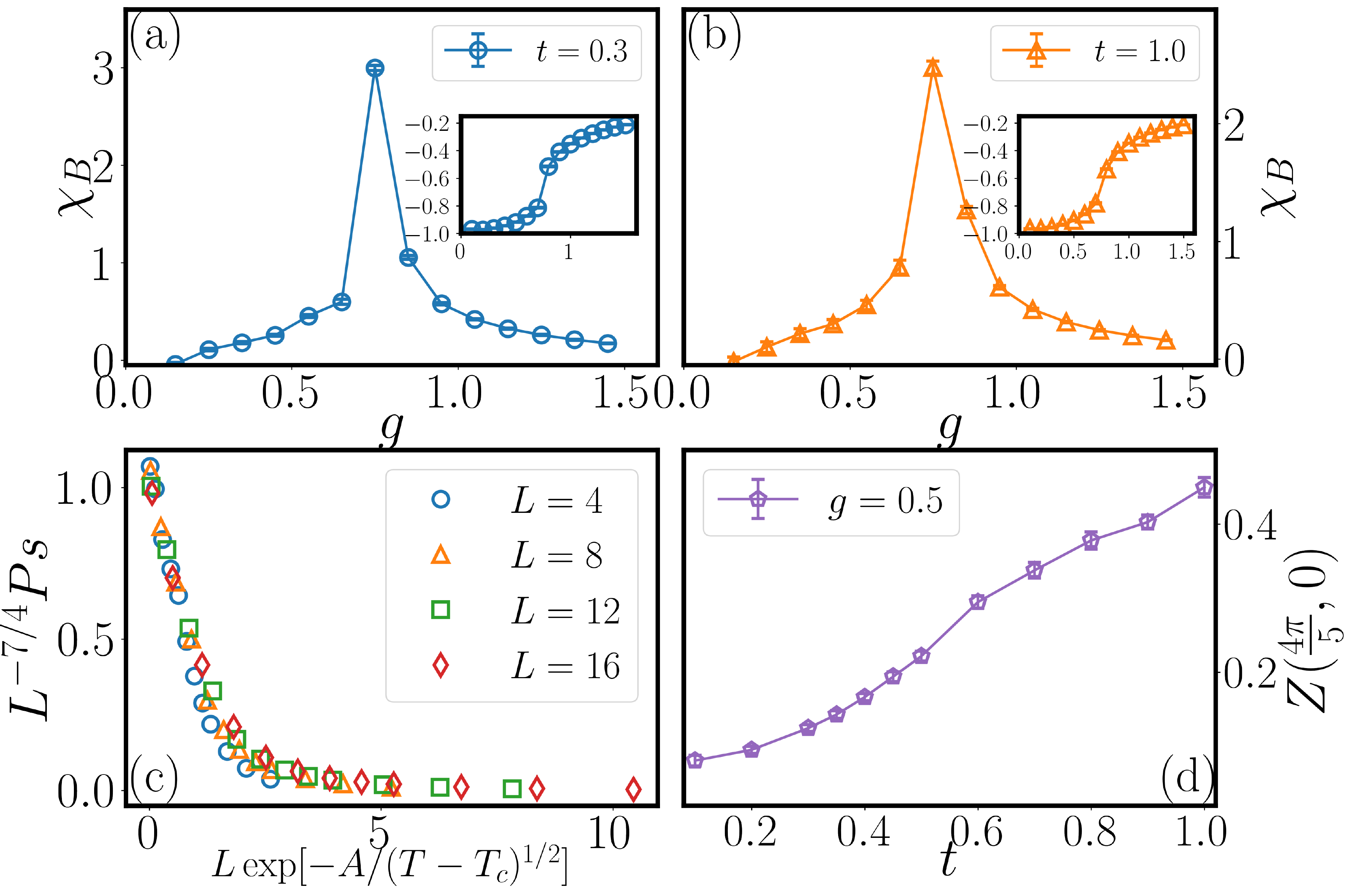}
\caption{(a) $Z_2$ flux susceptibility at $t=0.3, L=20, T=0.1$. By tuning $g$ the model goes from Fermi arc to confined FL. The peak in $\partial \langle B \rangle/\partial g$ denotes the transition point. (b) Similar measurement from deconfined FL to confined FL at $t=1.0, L=20, T=0.1$. (c) Data collapse of the s-wave pairing susceptibility $P_s$ at $t=0.3, g=1.4$ [solid circle in Fig.~\ref{fig:fig2}(a)]. The collapse signifies the KT transition $L^{2-\eta} P_{s}\sim L\cdot \exp[-\frac{A}{(T-T_{c})^{1/2}}]$ with $\eta=1/4$. $T_{c}=0.12$ gives the best collapse. (d) $Z(k=(\frac{4\pi}{5}, 0))$ at the antinodal point with $g=0.5, L=20, T=0.1$, through the transition from Fermi arc to the deconfined FL. In the former, the antinodal direction is gapped with small weight, and in the latter the large FS is formed with substantial weight; the transition point is $t \sim 0.5$.}
\label{fig:fig3}
\end{center}
\end{figure}

The confined FL phase appears with the enhancement of $g$ in $H_g$. Here the $Z_2$ gauge field is treated as a Higgsed field, and the phase corresponds to the normal metal phase in our previous orthogonal metal work~\cite{ChuangChen2020}. The transition from Fermi arc phase and deconfined FL phase to the confined FL phase can be seen from the the average $Z_2$ flux per plaquette, $B=\frac{1}{N}\sum_{\square}\prod_{b\in\square}\sigma^{z}_{b}$, and its susceptibility, $\partial \langle B \rangle/\partial g$, which were used to detect the confinement transition~\cite{Gazit2016,Gazit2018,Gazit2019}. Figures.~\ref{fig:fig3}(a) and ~\ref{fig:fig3}(b) show the results in sample paths as $g$ increases. There exist a change in $\langle B \rangle$ and a peak in $\partial \langle B \rangle/\partial g$ for $t=0.3$ in Fig.~\ref{fig:fig3}(a), and for $t=1.0$ in Fig.~\ref{fig:fig3}(b). These results signify the transitions from Fermi arc with $Z_2$ deconfinement to the confined FL at $g_c \sim 0.75$ and from deconfined FL to the confined FL at $g_c\sim 0.75$. The corresponding phase boundary in Fig.~\ref{fig:fig2}(a) is drawn in this way.

\begin{figure*}[htp!]
	\begin{center}
		\includegraphics[width=0.99\textwidth]{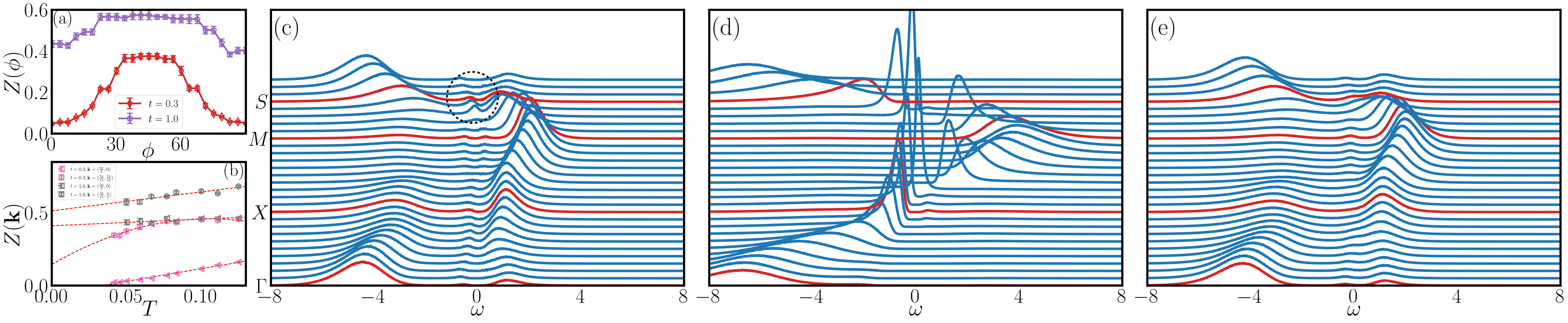}
		\caption{(a) Angle dependence of $Z(\phi)$ inside the Fermi arc and the deconfined FL phases. The angle $\phi = \arctan{\frac{k_y}{k_x}}$. (b) Temperature dependence of the quasiparticle fraction at nodal points $Z(\mathbf{k}=(\frac{7\pi}{12}, \frac{\pi}{2}))$ and $Z(\mathbf{k}=(\frac{7\pi}{12}, \frac{7\pi}{12}))$ and antinodal point $Z(\mathbf{k}=(\frac{5\pi}{6}, 0))$ inside the Fermi arc and deconfined FL phases. The parameters are the same as in Figs.~\ref{fig:fig1}(a) and ~\ref{fig:fig1}(d), respectively. (c) and (d) $A(\mathbf{k}, \omega)$ of the $c$ fermion, with $L=20, T=0.1$ inside the Fermi arc ($t=0.3, g=0.5$) and the confined FL ($t=1.0, g=1.4$) phases along the high-symmetry path in the BZ. The dashed circle in (c) denotes the discontinuous dispersion, signified by a spectral peak near $\omega=0$ close to the nodal points (from $\Gamma$ to $S$), which is consistent with the pseudogap behavior. (e) $A(\mathbf{k}, \omega)$ of the $c$ fermion at high temperature, with $L=20, T=0.2$, at the same parameters as in (c) ($t=0.3, g=0.5$), the pseudogap feature disappears due to thermal fluctuations.}
		\label{fig:fig4}
	\end{center}
\end{figure*}

It is interesting to note that we find that the confined FL is unstable towards $s$-wave pairing of $c$  fermions; we see this instability from the corresponding pairing susceptibility, $P_{s}=\frac{1}{L^2} \int_{0}^{\beta} d\tau \left \langle \Delta(\tau)\Delta^{\dagger}(0)+\text{H.c.} \right \rangle$ with $\Delta^{\dagger}_i = c^{\dagger}_{i\uparrow}c^{\dagger}_{i\downarrow}$. The finite-size collapse of $P_s$ with the exponent of the Kosterlitz-Thouless (KT) transition determines the $T_c$ ($T_{\text{KT}}$) as $P_s=L^{2-\eta} f\left[L \cdot \exp(-\frac{A}{(T-T_c)^{1/2}})\right]$ for $T \rightarrow T_c$ with $\eta=1/4$~\cite{isakov2003interplay,costa2018phonon}. The representative collapse is shown in Fig.~\ref{fig:fig3} (c), and $A=0.4$, $T_c = 0.12$ are obtained. In a similar manner, we performed the $P_s$ collapse in the phase diagram and found that $T_c$ reduce quickly as $t$ increases, the region of noticeable superconductivity is denoted by the blue shading in Fig.~\ref{fig:fig2}(a). Details of the finite-size analysis of $P_s$ are shown in Appendix~\ref{app:FS}.

Figure.~\ref{fig:fig3}(d) locates the Fermi arc to the deconfined FL transition, via the quasiparticle fraction $Z(\mathbf{k}) \sim \beta G(\mathbf{k},\beta/2)$ and we chose the antinodal point $\mathbf{k}=(\frac{4\pi}{5},0)$ with $g=0.5$ as a function of $t$. It is clear that inside the Fermi arc phase, $Z(\mathbf{k})$ is vanishingly small at this finite size ($L=20$), while when $t\sim 0.5$ there is a change in the slope of $Z(\mathbf{k})$ increase at the antinodal point suggesting the formation of a large FS, although the topological order still persists as shown in Fig.~\ref{fig:fig3}(b). The transition between the Fermi arc and the deconfined FL is estimated in this way.

\section{Fermi arc and pseudogap spectra}
Finally, we go back to the Fermi arc phase and clarify a few important points. The first one is whether the FS is indeed broken close to the zone boundary, and this can be confirmed from the comparison of the quasiparticle fraction along the nodal and antinodal directions. As shown in Fig.~\ref{fig:fig4}(a), it is clear that at finite temperature, inside the Fermi arc phase, the quasiparticle fraction along the antinodal direction is vanishingly small compared with that along the nodal direction, suggesting the existence of a pseudogap. In contrast, inside the deconfined FL, the quasiparticle fraction along both directions is finite. Another question is about the properties at the ground state. As shown in Fig.~\ref{fig:fig4}(b), as the temperature decreases, the finite quasiparticle fraction along the nodal direction inside the Fermi arc slowly extrapolates to a small but finite value, suggesting that the true ground state of the Fermi arc phase indeed has a finite density of states (we stress that the finite-size extrapolation has been applied to system sizes $L=\beta=24$; see Appendix~\ref{app:FS} for details). 

To make a closer comparison with the cuprate phenomenology~\cite{HongDing1996,Loeser1996,Marshall1996,Norman1998,Hashimoto2014}, we further compute the real-frequency single-particle spectra $A(\mathbf{k},\omega)$ of $c$ fermions in the Fermi arc and confined FL phases. The spectra are obtained from stochastically analytic continuation of the imaginary time Green's function from the QMC simulation. Such a methodology has been successfully employed in various strongly correlated systems~\cite{Sandvik1998,Beach2004,Constrained2016Sandvik,Nearly2017Shao,Dynamical2018Sun,NSMA2018,CJHuang2018,NVMa2019,ZhengYan2020,BKT2019,BKT2020,CKZhou2020}. The obtained spectra are shown in Fig.~\ref{fig:fig4}(c) and ~\ref{fig:fig4}(d); it is interesting to see that along the high-symmetry-path, the Fermi arc phase [Fig.~\ref{fig:fig4}(c)] has a discontinuous dispersion, signified by a spectral peak near $\omega=0$ close to the nodal points [from $\Gamma$ to S, indicated by the dashed circle in Fig.~\ref{fig:fig4}(c)] and its disappearance along the other parts of the path (especially along the antinodal direction X-$\Gamma$). This is consistent with the pseudogap and Fermi arc phenomena, i.e., the breaking of the FS and the violation of Luttinger's theorem, and is in sharp contrast with the continuous dispersion of the confined FL phase [Fig.~\ref{fig:fig4}(d)], where the quasiparticle peaks are pronounced at all the momenta, in particular, close to $\omega=0$. We also find, as the temperature rises (with $T=0.2$), the pseudogap near X disappears due to the thermal fluctuations, as shown in Fig.~\ref{fig:fig4}(e). 


\section{Discussions}
By doping the orthogonal metal, we reveal a Fermi arc and pseudogap state at generic filling in a lattice model of correlate electrons with unbiased QMC simulations. Our observations share a phenomenological similarity with the cuprate  experiments~\cite{HongDing1996,Loeser1996,Marshall1996,Norman1998,Kanigel2006,Kondo2013,Hashimoto2014}: There is a strong depletion in the quasiparticle weight at antinodal points; there is no translational symmetry breaking and the state appears to violate LT; and the large and closed FS emerges as the hopping of gauge-neutral $c$ fermions increases and superconductivity therein. In a more general sense, our results therefore provide a concrete model realization and theoretical framework for the coupling between gauge field and fermions, and in the process, generate the rich phenomena of the pseudogap, the Fermi arc, and superconductivity in generic correlated electron systems. The deeper connection of our theoretical model with that of the Hubbard-type model, where the emergent gauge field coupled to fractionalized quasiparticles is expected~\cite{Wen1996,DHKim1997,PLee2006,Osborne2020}, and experimental reality is ready to be explored.

\acknowledgements
We thank S. Bhattacharjee, S. Gazit, F. Assaad, M. Metlitski, T. Senthil and S. Sachdev for helpful discussions. C.C. and Z.Y.M acknowledge support from the RGC of Hong Kong SAR of China (Grants No. 17303019 and No. 17301420), MOST through the National Key Research and Development Program (Grant No. 2016YFA0300502), and the Strategic Priority Research Program of the Chinese Academy of Sciences (Grant No. XDB33020300). T.Y. and Y.Q. acknowledge support from MOST under Grant No. 2015CB921700 and from NSFC under Grant No. 11874115. We thank the Center for Quantum Simulation Sciences in the Institute of Physics, Chinese Academy of Sciences, the Computational Initiative at the Faculty of Science and the Information Technology Services at the University of Hong Kong and the Tianhe-1A, Tianhe-2, and Tianhe-3 prototype platforms at the National Supercomputer Centers in Tianjin and Guangzhou for their technical support and generous allocation of CPU time.

\appendix

\section{Quantum Monte Carlo Implementation}
\label{app:QMC}
In this appendix, we discuss how the quantum Monte Carlo simulation of the model is implemented; while part of this introduction is given in our previous orthogonal metal work~\cite{ChuangChen2020}, the addition of the composite fermion hopping term $H_c$ has greatly increased the complexity of Monte Carlo simulations, and we have managed to maintain a similar level of the numerical stability with the block update scheme. We will first recap the construction of the partition function and then pay more attention to the update scheme of the $H_c$ term.

After discretizing the imaginary time $\beta=\Delta\tau L_{\tau}$ and performing the trace of the Ising matter field in the $S^{z}$ basis, the trace of the $Z_2$ gauge field in the $\sigma^{z}$ basis, and the trace of fermion degrees of freedom to obtain the fermion determinant, the partition function of our model can be written as
\begin{widetext}
\begin{align}
Z= & \text{Tr}\left\{ e^{-\beta H} \right\} \nonumber \\
= & \sum_{\{S^{z}_{i},\sigma^{z}_{b}\}}\exp\left[\sum_{l,\langle i,j\rangle}\Delta\tau JS_{i}^{z}(l)\sigma_{b}^{z}(l)S_{j}^{z}(l)+\sum_{i,\langle l,l'\rangle}\gamma_{s}S_{i}^{z}(l)S_{i}^{z}(l')\right]\times \exp\left[\sum_{l,\square}\Delta\tau K\prod_{b\in\square}\sigma_{b}^{z}(l)+\sum_{b,\langle l,l'\rangle}\gamma_{\sigma}\sigma_{b}^{z}(l)\sigma_{b}^{z}(l')\right]\times \nonumber \\
& \left|\det\left(I+\prod_{l=L_{\tau}}^{1}\mathbf{B}(l)\right)\right|^{2},
\label{eq:eq2}
\end{align}
\end{widetext}
where $\gamma_{s}=-\frac{1}{2}\ln\left[\tanh(\Delta\tau h)\right]$,
$\gamma_{\sigma}=-\frac{1}{2}\ln\left[\tanh(\Delta\tau g)\right]$,
and matrices $\mathbf{B}(l)=\exp\left[V(l)\right]$ with $V(l)$ (imaginary time-slice index $l$ takes values $1,\cdots, L_\tau$; the spatial site indices $i,j$ take the values $1,\cdots,L^2$) having elements $V(l)_{\langle i,j\rangle}=\Delta\tau t\sigma_{b}^{z}(l)$
and $V(l)_{i,i}=\Delta\tau\mu$. We will leave the discussion of the $H_c$ term inside $\mathbf{B}(l)$ to Appendix~\ref{app:QMC}. The square outside of the determinant
comes from two species of fermion (spin up and down). As the bosonic
parts of weights are always positive and the fermion part of the weight
is a square of the determinant of the real matrix, the whole weight will always be semipositive, and it is absent of a sign problem.

We therefore use the determinant quantum Monte Carlo (DQMC) approach to simulate this model,
which has been widely used in simulating fermion boson coupled lattice
models, and more details can be found in the recent review in Ref.~\cite{XiaoYanXu2019}. The local updates
are performed on the Ising matter field $\{S_{i}^{z}\}$ and $Z_2$
gauge fields $\{\sigma_{b}^{z}\}$ in a space-time configurational
space with volume $L\times L \times L_{\tau}$, where $L_{\tau}=\beta/\Delta\tau$
with $\Delta\tau=0.1$ and $\beta=L=$12, 14, ..., 20, 24.

\begin{figure}[htp!]
\begin{center}
\includegraphics[width=0.9\columnwidth]{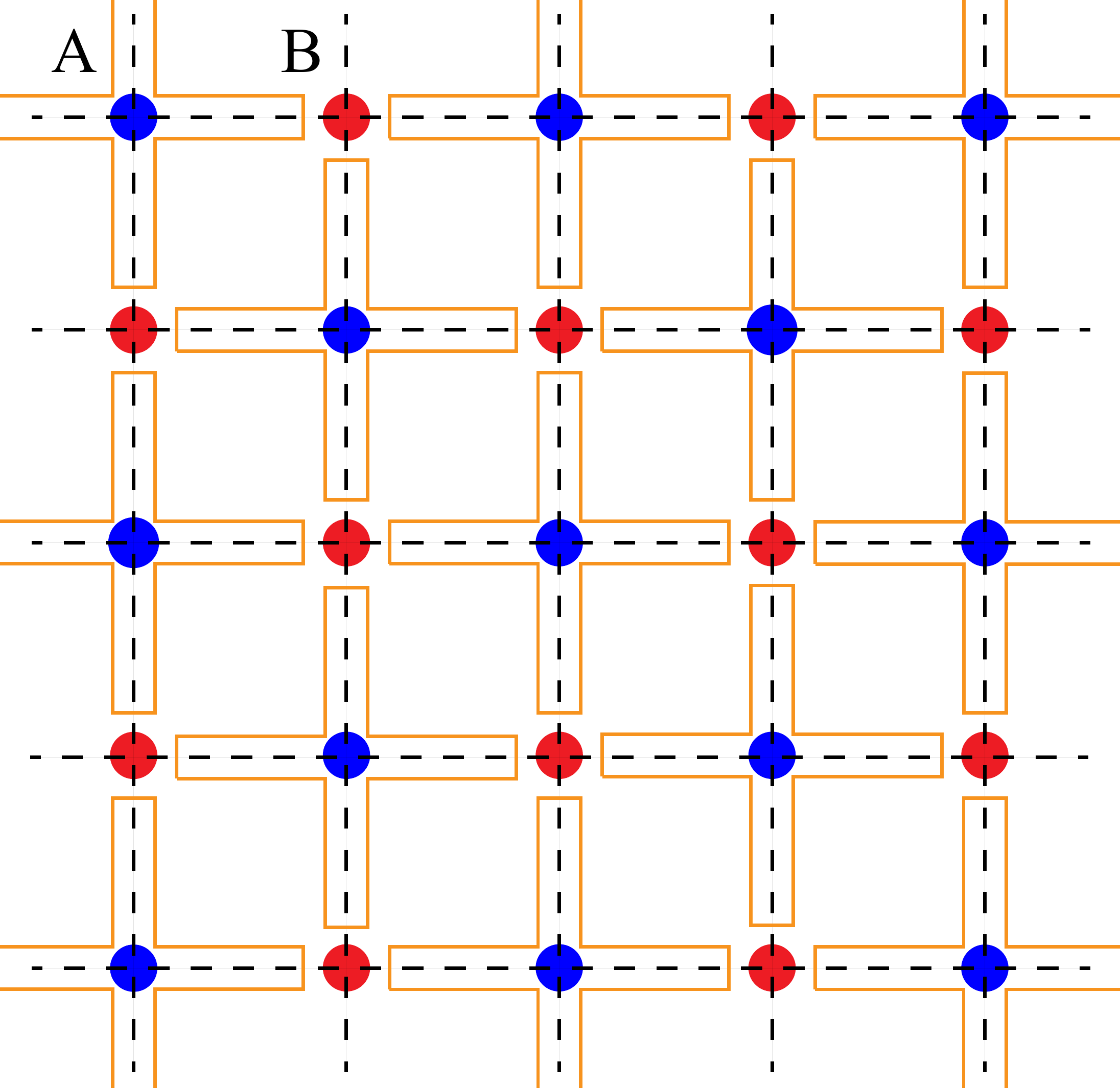}
\caption{Blue and red dots stand for the A and B sublattices of $S^{z}_{i}$, respectively, and the crosses originating from the blue and red dots are the four nearest neighbor (NN) interactions of $S^{z}_{i}S^{z}_{j}$.}
\label{fig:fig5}
\end{center}
\end{figure}

\subsection{Scheme to update $H_c$ term}
$H_c$ is the hopping of the composite $c$-fermion as a combination of the orthogonal fermion $f_{i,\alpha}$ and spin matter field $S_{i}^{z}$. After the path integral of the partition function, it is equivalent to view the spin variable $S_{i}^{z}=\pm 1$ entering the hopping matrix of the $f $fermion.

\begin{equation}
H_{c} = -t \sum_{\langle i,j\rangle} f^{\dagger}_{i,\alpha}S^{z}_{i}f_{j,\alpha}S^{z}_{j} + \text{H.c.}
\end{equation}

Using Trotter decomposition, we can write the $\mathbf{B}$ matrix in the fermion determinant in the following form

\begin{equation}
\mathbf{B}^{\tau} = e^{-\Delta \tau \mathbf{T}_{\sigma, \tau}} \cdot e^{-\Delta \tau \mathbf{T}_{\mu, \tau}} \cdot e^{-\Delta \tau \mathbf{T}_{S^z, \tau}}
\end{equation}
where $\mathbf{T}_{\sigma}$ is the matrix from the $H_{f}$ term and $\mathbf{T}_{\mu}$ is the chemical potential matrix. $\mathbf{T}_{S^z}$ is the matrix $\mathbf{T}_{S^z, ij}=S_{i}^{z}S_{j}^{z}$. As shown in Fig.~\ref{fig:fig5}, we can further exploit the Trotter decomposition to split the $\mathbf{T}_{S^z}$ matrix into A-B sublattice form
\begin{widetext}
\begin{align}
e^{-\Delta \tau \mathbf{T}_{S^z, \tau}} =& e^{-\Delta \tau \mathbf{T}_{S^z, A_1,\tau}} \cdot e^{-\Delta \tau \mathbf{T}_{S^z, A_2,\tau}} \cdots e^{-\Delta \tau	 \mathbf{T}_{S^z, A_{N/2}, \tau}} + \mathcal{O}\left( \Delta	\tau^2 \right) \label{eq:eqa}\\
=& e^{-\Delta \tau	 \mathbf{T}_{S^z, B_1, \tau}} \cdot e^{-\Delta \tau \mathbf{T}_{S^z, B_2, \tau}} \cdots e^{-\Delta \tau	 \mathbf{T}_{S^z, B_{N/2}, \tau}} + \mathcal{O}\left( \Delta	\tau^2 \right)
\label{eq:eqb}
\end{align}
\end{widetext}
where $N=L^2$ is the number of sites and $A$ and $B$ stand for the elements between $A_i$ or $B_i$ site and its four neighboring sites, respectively. Matrix $\mathbf{T}_{S^z, A_i/B_i}$ is zero except for the entries connected by site $A_i$ or $B_i$ and its four neighboring sites, illustrated in Fig.~\ref{fig:fig5} for the Eq.(~\ref{eq:eqa}) type decomposition.

Unlike the DQMC approach for the Hubbard model, where in order to calculate the ratio of determinants and update the Green's function only one element of the Hubbard-Stratonovich (HS) field matrix is involved, we have four elements that are changed when we update one $S_i^z$ in the $c$-fermion hopping term $H_c$. Now we discuss how to calculate the ratio and update the Green's function with multiple changing matrix elements. Firstly, we introduce the $\mathbf{\Delta}$ matrix

\begin{eqnarray}
e^{-\Delta \tau \mathbf{T}_{S^z, A_i', \tau}} &=& \left( \mathbf{1} + \mathbf{\Delta} \right) e^{-\Delta \tau \mathbf{T}_{S^z, A_i, \tau}}, \nonumber\\
e^{-\Delta \tau \cdot 2 \cdot \mathbf{T}_{S^z, A_i, \tau}} &=& \left( \mathbf{1} + \mathbf{\Delta} \right)
\end{eqnarray}
Once we propose an update $S_i^z \rightarrow -S_i^z$,  $\mathbf{T}_{S^z, A_i', \tau}  = -\mathbf{T}_{S^z, A_i', \tau}$. One lattice site has four nearest-neighbor hoppings, so we have a total of $2^4=16$ $\mathbf{\Delta}$ matrices. We can compute all of them in advance to avoid repeatedly calculating them during the simulation.

Below is the general scheme to calculate the ratio and update the Green's function with the $k$-dimensional $\mathbf{\Delta}$ matrix ~\cite{xu_thesis}.

Define
\begin{equation}
\mathbf{B}^{M}\cdots\mathbf{B}^{\tau+1}\equiv\mathbf{B}(\beta,\tau)
\end{equation}
\begin{equation}
\mathbf{B}^{\tau}\cdots\mathbf{B}^{1}\equiv\mathbf{B}(\tau,0)
\end{equation}

Try to flip $s_{i,\tau}$ ,
\begin{equation}
\det\left(\mathbf{1}+\mathbf{B}(\beta,\tau)\mathbf{B}(\tau,0)\right)\rightarrow\det\left(\mathbf{1}+\mathbf{B}(\beta,\tau)(\mathbf{1}+\boldsymbol{\Delta})(\mathbf{B}(\tau,0)\right)
\end{equation}

The weight ratio is
\begin{align}
 & \frac{\det\left(\mathbf{1}+\mathbf{B}(\beta,\tau)(\mathbf{1}+\boldsymbol{\Delta})(\mathbf{B}(\tau,0)\right)}{\det\left(\mathbf{1}+\mathbf{B}(\beta,\tau)\mathbf{B}(\tau,0)\right)} \nonumber \\
= & \det\left[\mathbf{1}+\boldsymbol{\Delta}\left(\mathbf{1}-(\mathbf{1}+\mathbf{B}(\tau,0)\mathbf{B}(\beta,\tau))^{-1}\right)\right] \nonumber \\
= & \det\left[\mathbf{1}+\boldsymbol{\Delta}(\mathbf{1}-\mathbf{G}(\tau,\tau))\right]
\end{align}

If the update is accepted, we also need to update the Green's function
\begin{widetext}
\begin{eqnarray}
\mathbf{G}'(\tau,\tau) & = & \left[\mathbf{1}+\left(\mathbf{1}+\boldsymbol{\Delta}\right)\mathbf{B}(\tau,0)\mathbf{B}(\beta,\tau)\right]^{-1} \nonumber \\
 & = & \left[\mathbf{1}+\mathbf{B}(\tau,0)\mathbf{B}(\beta,\tau)\right]^{-1}\left[\left(\mathbf{1}+\left(\mathbf{1}+\boldsymbol{\Delta}\right)\mathbf{B}(\tau,0)\mathbf{B}(\beta,\tau)\right)\left(\left(\mathbf{1}+\mathbf{B}(\tau,0)\mathbf{B}(\beta,\tau)\right)^{-1}\right)\right]^{-1}
\end{eqnarray}
\end{widetext}

As we have $\mathbf{G}\equiv\mathbf{G}(\tau,\tau)=\left[\mathbf{1}+\mathbf{B}(\tau,0)\mathbf{B}(\beta,\tau)\right]^{-1}$,
we also denote $\mathbf{A}\equiv\mathbf{B}(\tau,0)\mathbf{B}(\beta,\tau)\equiv\mathbf{G}^{-1}-\mathbf{1}$,
then we have
\begin{eqnarray}
\mathbf{G}'(\tau,\tau) & = & \mathbf{G}\left[\left(\mathbf{1}+\left(\mathbf{1}+\boldsymbol{\Delta}\right)\mathbf{A}\right)\mathbf{G}\right]^{-1} \nonumber \\
 & = & \mathbf{G}\left[\left(\mathbf{1}+\left(\mathbf{1}+\boldsymbol{\Delta}\right)\left(\mathbf{G}^{-1}-\mathbf{1}\right)\right)\mathbf{G}\right]^{-1} \nonumber \\
 & = & \mathbf{G}\left[\mathbf{1}+\boldsymbol{\Delta}\left(\mathbf{1}-\mathbf{G}\right)\right]^{-1}
\end{eqnarray}

Note that $\boldsymbol{\Delta}\left(\mathbf{1}-\mathbf{G}\right)$ only
has $k$ rows that are nonezero; thus $\boldsymbol{\Delta}\left(\mathbf{1}-\mathbf{G}\right)$ can be formulated as the cross
product of two rectangular matrix, $\boldsymbol{\Delta}\left(\mathbf{1}-\mathbf{G}\right)\equiv\mathbf{U}\mathbf{V}$,
with
\begin{widetext}
\begin{equation}
\mathbf{\mathbf{U}}=\left[\begin{array}{ccc}
0 & 0 & \cdots\\
\vdots & \vdots & \cdots\\
\Delta_{ii} & \Delta_{ij} & \cdots\\
\vdots & \vdots & \cdots\\
\Delta_{ji} & \Delta_{jj} & \cdots\\
0 & 0 & \cdots\\
\vdots & \vdots & \cdots\\
0 & 0 & \cdots
\end{array}\right]_{N\times k}
\end{equation}
and
\begin{equation}
\mathbf{V}=-\left[\begin{array}{cccccccc}
G_{i1} & \cdots & G_{ii}-1 & \cdots & G_{ij} & \cdots & \cdots & G_{i,N}\\
G_{j1} & \cdots & G_{ji} & \cdots & G_{jj}-1 & \cdots & \cdots & G_{j,N}\\
\vdots & \vdots & \vdots & \vdots & \vdots & \vdots & \vdots & \vdots
\end{array}\right]_{k\times N}
\end{equation}
\end{widetext}

Then with the help of the generalized Sherman-Morrison formula $\left(\mathbf{I}+\mathbf{U}\mathbf{V}\right)^{-1}=\mathbf{I}-\mathbf{U}\left(\mathbf{I}_{k}+\mathbf{V}\mathbf{U}\right){}^{-1}\mathbf{V}$,
we have
\begin{eqnarray}
\mathbf{G}'(\tau,\tau) & = & \mathbf{G}\left[\mathbf{1}+\boldsymbol{\Delta}\left(\mathbf{1}-\mathbf{G}\right)\right]^{-1} \nonumber \\
 & = & \mathbf{G}\left(\mathbf{1}+\mathbf{U}\mathbf{V}\right)^{-1}\nonumber \\
 & = & \mathbf{G}\left[\mathbf{I}-\mathbf{U}\left(\mathbf{I}_{k}+\mathbf{V}\mathbf{U})^{-1}\mathbf{V}\right)\right]\nonumber \\
 & = & \mathbf{G}-\mathbf{G}\mathbf{U}\left(\mathbf{I}_{k}+\mathbf{V}\mathbf{U}\right){}^{-1}\mathbf{V}
\end{eqnarray}

Now we try to formulate it in a more standard form (an easy extension to delay the update).
We can factorize $\mathbf{U}$ as
\begin{widetext}
\begin{equation}
\mathbf{U}=\left[\begin{array}{ccc}
0 & 0 & \cdots\\
\vdots & \vdots & \cdots\\
\Delta_{ii} & \Delta_{ij} & \cdots\\
\vdots & \vdots & \cdots\\
\Delta_{ji} & \Delta_{jj} & \cdots\\
0 & 0 & \cdots\\
\vdots & \vdots & \cdots\\
0 & 0 & \cdots
\end{array}\right]_{N\times k}=\left[\begin{array}{ccc}
0 & 0 & \cdots\\
\vdots & \vdots & \cdots\\
1_{ii} & 0 & \cdots\\
\vdots & \vdots & \cdots\\
0 & 1_{jj} & \cdots\\
0 & 0 & \cdots\\
\vdots & \vdots & \cdots\\
0 & 0 & \cdots
\end{array}\right]_{N\times k}\left[\begin{array}{ccc}
\Delta_{ii} & \Delta_{ij} & \cdots\\
\Delta_{ji} & \Delta_{jj} & \cdots\\
\vdots & \vdots & \ddots
\end{array}\right]_{k\times k}\equiv\tilde{\mathbf{U}}\mathbf{D}
\end{equation}
\end{widetext}

Redefine $\mathbb{U}=\mathbf{G}\tilde{\mathbf{U}}$, $\mathbb{S}\equiv\mathbf{D}\left(\mathbf{I}_{k}+\mathbf{V}\mathbf{U}\right){}^{-1}$, and $\mathbb{V}=-\mathbf{V}$ with

\begin{equation}
\mathbb{U}\equiv\mathbf{G}\tilde{\mathbf{U}}=\left[\begin{array}{ccc}
G_{1i} & G_{2j} & \cdots\\
\vdots & \vdots & \cdots\\
G_{ii} & G_{ij} & \cdots\\
\vdots & \vdots & \cdots\\
G_{ji} & G_{jj} & \cdots\\
\vdots & \vdots & \cdots\\
\vdots & \vdots & \cdots\\
G_{Ni} & G_{Nj} & \cdots
\end{array}\right]_{N\times k}
\end{equation}

\begin{widetext}
\begin{eqnarray}
\mathbb{S} & \equiv & \mathbf{D}\left(\mathbf{I}_{k}+\mathbf{V}\mathbf{U}\right){}^{-1}\\
 & = & \left[\begin{array}{ccc}
\Delta_{ii} & \Delta_{ij} & \cdots\\
\Delta_{ji} & \Delta_{jj} & \cdots\\
\vdots & \vdots & \ddots
\end{array}\right]_{k\times k}\left(\mathbf{I}_{k}-\left[\begin{array}{cccccccc}
G_{i1} & \cdots & G_{ii}-1 & \cdots & G_{ij} & \cdots & \cdots & G_{i,N}\\
G_{j1} & \cdots & G_{ji} & \cdots & G_{jj}-1 & \cdots & \cdots & G_{j,N}\\
\vdots & \vdots & \vdots & \vdots & \vdots & \vdots & \vdots & \vdots
\end{array}\right]_{k\times N}\left[\begin{array}{ccc}
0 & 0 & \cdots\\
\vdots & \vdots & \cdots\\
\Delta_{ii} & \Delta_{ij} & \cdots\\
\vdots & \vdots & \cdots\\
\Delta_{ji} & \Delta_{jj} & \cdots\\
0 & 0 & \cdots\\
\vdots & \vdots & \cdots\\
0 & 0 & \cdots
\end{array}\right]_{N\times k}\right)^{-1}
\end{eqnarray}
and
\begin{equation}
\mathbb{V}\equiv-\mathbf{V}=\left[\begin{array}{cccccccc}
G_{i1} & \cdots & G_{ii}-1 & \cdots & G_{ij} & \cdots & \cdots & G_{i,N}\\
G_{j1} & \cdots & G_{ji} & \cdots & G_{jj}-1 & \cdots & \cdots & G_{j,N}\\
\vdots & \vdots & \vdots & \vdots & \vdots & \vdots & \vdots & \vdots
\end{array}\right]_{k\times N}
\end{equation}
\end{widetext}
Then we have the weight ratio $r=\det(\mathbf{I}_{k}+\mathbf{V}\mathbf{U})$
and
\begin{equation}
\mathbf{G}'(\tau,\tau)=\mathbf{G}+\mathbb{U}\mathbb{S}\mathbb{V}
\end{equation}

Back to our partition function in Eq.~\eqref{eq:eq2} and the update of the $H_c$ term, as discussed in the beginning of this section, we have $k=5$ for updating $S_i^z$. Because we use either Eq.~\eqref{eq:eqa} or Eq.~\eqref{eq:eqb} to calculate the determinant, we can only update half of $S_i^z$ in one  Monte Carlo sweep (in the usual sense). So our scheme is to perform one sweep to update the A sublattice with the Green's function calculated using Eq.~\eqref{eq:eqa} and recalculate the Green's function using Eq.~\eqref{eq:eqb}, and then we update all B sublattice sites. One "sweep" therefore contains two usual sweeps.

\section{MEAN FIELD CALCULATION OF THE DOPED OSM PHASE}
\label{app:MF}
In this appendix, we present a mean-field calculation of the spectral properties in the Fermi arc phase of our phase diagram. The calculation here is an extension to the doping case of the calculation for the orthogonal semi-metal case in Ref.~\cite{Gazit2019}.

We consider the limit of $g=t=0$ and neglect gauge field fluctuations and $c$-fermion hopping terms. Since the $c$-fermion spectral function we calculate is a gauge-invariant quantity, we can choose a gauge condition which is $\sigma_{r,\hat{x}}=(-1)^{r_y}$ and $\sigma_{r,\hat{y}}=1$. The $f$-fermion ($\tau^z$ field) is a free fermion (scalar field) hopping in the background of the static gauge field. So we take the mean-field Hamiltonian
\begin{align}
&\mathcal{H}^{\text{MF}}_f=-\sum_{r,\eta}t_{r,\eta}f^\dagger_{r,\alpha}f_{r+\eta,\alpha}-\mu\sum_r f^\dagger_{r,\alpha}f_{r,\alpha}\\
&\mathcal{H}^{\text{MF}}_\phi=\sum_r \frac{1}{2}\pi^2_r+\frac{1}{2}\omega^2\left(\sum_r\Delta\phi_r^2+\frac{1}{2}\sum_{r,\eta}(\phi_r-t_{r,\eta}\phi_{r+\eta})^2\right)
\end{align}
where $\phi_r$ is a real scalar field and $\pi_r$ is its canonical momentum. $\eta$ takes a value in $\{\pm\hat{x},\pm\hat{y}\}$ and the hopping amplitude $t_{r,\eta}=(-1)^{r_y}\delta_{\eta,\pm\hat{x}}+\delta_{\eta,\pm\hat{y}}$.

The gauge condition breaks the translation symmetry, so momentum space of the mean-field Hamiltonian is defined on a reduced Brillouin zone($0<k_x<2\pi,0<k_y<\pi$). For the $f$ fermion, substituting $f_{r,\alpha}=\frac{1}{\sqrt{N}}\sum_k f_{k,\alpha} e^{ikr}$
\begin{widetext}
\begin{equation}
\begin{aligned}
H^{\text{MF}}_f&=-\sum_{k,k'}f^\dagger_{k,\alpha}f_{k',\alpha}\left(\sum_\eta e^{ik'\eta}\left(\delta_{\eta,\pm\hat{x}}\delta_{k,k'+\pi\hat{k_y}}+\delta_{\eta,\pm\hat{y}}\delta_{k,k'}\right)+\mu\delta_{k,k'}\right)\\
&=-\sum_{k}2\cos(k_x)f^\dagger_{k,\alpha}f_{k-\pi\hat{k_y},\alpha}+\left(2\cos(k_y)+\mu\right)f^\dagger_{k,\alpha}f_{k,\alpha}\\
&=-\sideset{}{'}\sum_{k}\begin{pmatrix} f^\dagger_{0,\alpha}(k)&f^\dagger_{\pi,\alpha}(k) \end{pmatrix}
\begin{pmatrix} 2\cos(k_y)+\mu&2\cos(k_x)\\2\cos(k_x)&-2\cos(k_y)+\mu \end{pmatrix}
\begin{pmatrix} f_{0,\alpha}(k)\\f_{\pi,\alpha}(k) \end{pmatrix}
\end{aligned}
\end{equation}
\end{widetext}
where $f_{0,\alpha}(k)=f_{k,\alpha}$, $f_{\pi,\alpha}(k)=f_{k+\pi\hat{k_y},\alpha}$ and $\sideset{}{'}\sum$ is the sum of momentum in the reduced Brillouin zone. Diagonalizing the Hamiltonian, we get the energy spectrum $\epsilon_\pm(k)$ and the eigen-modes $f_{\rho,\alpha}(k)=V_{\rho,\gamma}(k)f_{\gamma,\alpha}(k)$, where $\gamma=\pm$ , $\rho=0/\pi$ and $V_{\rho,\gamma}$ diagonalize the Hamiltonian. It is useful to represent $f_{k,\alpha}$ by $f_{\gamma,\alpha}(k)$
\begin{widetext}
\begin{equation}
f_{k,\alpha}=V_{\rho(k),\gamma}(P(k))f_{\gamma,\alpha}(P(k))\quad , \quad
\rho(k)=
\begin{cases}
0,\quad k_y\in[0,\pi)\\
\pi,\quad k_y\in[\pi,2\pi)
\end{cases} , \quad
P(k)=
\begin{cases}
k_y,\quad k_y\in[0,\pi)\\
k_y-\pi,\quad k_y\in[\pi,2\pi)
\end{cases}
\end{equation}
\end{widetext}
The $f$-fermion spectrum function is [it is equivalent to understand $A_f(k,k',\omega)$ as $A_f(k,\omega)_{\rho,\rho'}$; the momentum of the later is in the reduced BZ.]
\begin{widetext}
\begin{equation}
\begin{aligned}
A_f(k,k',\omega)&=\frac{1}{2\pi}\int dt e^{i\omega t} \left\langle \left\{ f_k(t) , f_{k'}^\dagger \right \} \right\rangle \\
&=\sum_\gamma \delta(\omega-\epsilon_\gamma(P(k))) \delta_{P(k),P(k')}V_{\rho(k),\gamma}(P(k))V_{\rho(k'),\gamma}(P(k))
\end{aligned}
\end{equation}
\end{widetext}

\begin{figure}[htp!]
	\begin{center}
		\includegraphics[width=0.9\columnwidth]{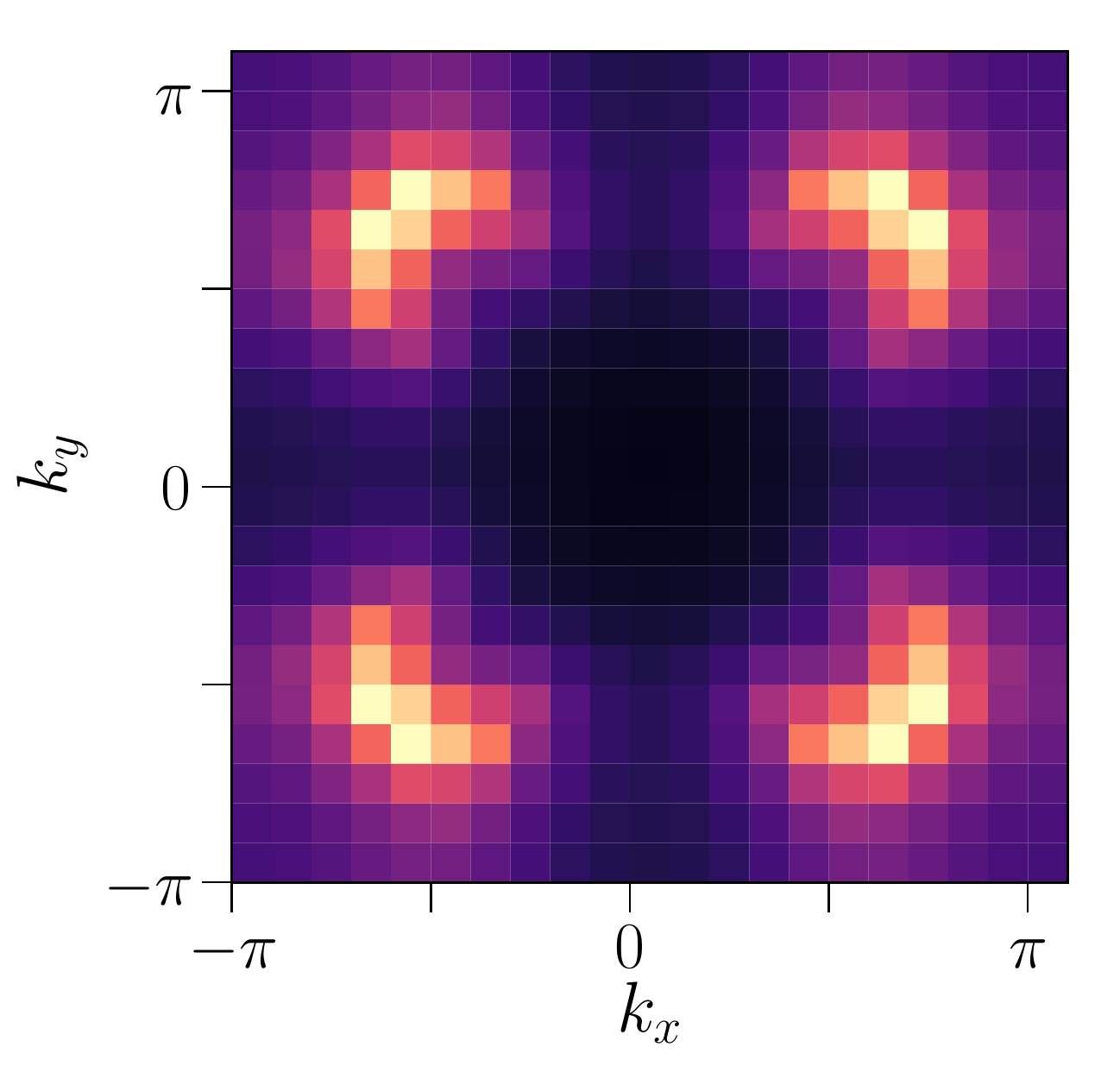}
		\caption{Mean-field $c$-fermion spectral function $A(\mathbf{k},\omega=0)$. The parameters are $L=20$, $T=0.1$, $\mu=1.2$, $\omega=1$ and $\Delta=-1.1$.}
		\label{fig:fig6}
	\end{center}
\end{figure}

For the scalar field, substituting $\phi_r=\frac{1}{\sqrt{N}}\sum_k \phi_r e^{ikr}$
\begin{widetext}
\begin{equation}
\begin{aligned}
\mathcal{H}^{\text{MF}}_\phi&=\sum_k \frac{1}{2}\pi_k\pi_{-k}+\frac{1}{2}m\omega^2\left(\Delta \phi_k\phi_{-k}
+(4-2\cos(k_y))\phi_k\phi_{-k}-(2\cos(k_x))\phi_k\phi_{-k-\pi\hat{k_y}} \right) \\
&=\sideset{}{'}\sum_{k,\rho}\frac{1}{2}\pi_{k,\rho}\pi_{-k,\rho}+\sideset{}{'}\sum_{k}\frac{1}{2}\omega^2
\begin{pmatrix} \phi_{0}(k)&\phi_{\pi}(k) \end{pmatrix}
\begin{pmatrix} \Delta+4-2\cos(k_y)&-2\cos(k_x)\\-2\cos(k_x)&\Delta+4+2\cos(k_y) \end{pmatrix}
\begin{pmatrix} \phi_{0}(-k)\\ \phi_{\pi}(-k) \end{pmatrix}
\end{aligned}
\end{equation}
\end{widetext}
Diagonalizing the frequency matrix, we get the eigenfrequency $\omega_\kappa(k)$ and the normal modes $\phi_{\rho}(k)=U_{\rho,\kappa}(k)\phi_{\kappa}(k)$ [caution: the eigenvalue of the matrix is the square of $\omega_\kappa(k)$, so we must set $\Delta>2\sqrt{2}-4$; otherwise the eigenvalue will be negative]. To diagonalize the Hamiltonian, we introduce the operators
\begin{equation}
\begin{cases}
a_\kappa(k)=\sqrt{\frac{1}{2}\omega_\kappa(k)}\phi_\kappa(k)+i\sqrt{\frac{1}{2\omega_\kappa(k)}}\pi_\kappa(k) \\
a^\dagger_\kappa(-k)=\sqrt{\frac{1}{2}\omega_\kappa(k)}\phi_\kappa(k)-i\sqrt{\frac{1}{2\omega_\kappa(k)}}\pi_\kappa(k)
\end{cases}
\end{equation}
then the Hamiltonian becomes
\begin{equation}
\mathcal{H}^{\text{MF}}_\phi=\sideset{}{'}\sum_{k,\kappa}\omega_\kappa(k)a^\dagger_\kappa(k)a_\kappa(k)+\text{const}.
\end{equation}
Representing $\phi_k$ by $a$ and $a^\dagger$,
\begin{equation}
\phi_k=\frac{U_{\rho(k),\kappa}(P(k))}{\sqrt{2\omega_\kappa(P(k))}}\left( a_\kappa(P(k))+a^\dagger_\kappa(-P(k)) \right)
\end{equation}

There is some subtlety in the commutation relation. Notice that $\phi_0(-k)$($\phi_\pi(-k)$) is defined as $\phi_{-k}$($\phi_{-k-\pi\hat{k_y}}$); we will get a strange commutation relation between positive $k_y$ and negative $k_y$. In the calculation of the Hamiltonian, we do not meet any problem because the commutation relations have the same momentum, but that is not the case in the calculation of the spectrum function. At least, we can require $k,k'>0$ to avoid the subtlety.
\begin{widetext}
\begin{equation}
\begin{aligned}
A_\phi(k,k',\omega)&=\frac{1}{2\pi}\int dt e^{i\omega t} \left\langle \left[ \phi_k(t) , \phi_{-k'} \right ] \right\rangle \\
&=\frac{1}{2\pi}\int dt e^{i\omega t} \sum_{\kappa,\kappa'}U_{\rho(k),\kappa}(P(k))U_{\rho(-k'),\kappa'}(P(-k'))\frac{1}{\sqrt{
4\omega_{\kappa}(P(k))\omega_{\kappa'}(P(-k'))  }}\\
&\quad\quad\quad\quad\quad\quad\times\left\langle  \left[
a_{\kappa}(P(k))(t)+a^\dagger_\kappa(-P(k))(t), a_{\kappa}(P(-k'))+a^\dagger_\kappa(-P(-k'))
\right ]  \right\rangle\\
&=\sum_{\kappa}\left[ \delta(\omega-\omega_\kappa(P(k)))-\delta(\omega+\omega_\kappa(P(k))) \right]\delta_{P(k),-P(-k')}\\
&\quad\quad\quad\quad\quad\quad\times U_{\rho(k),\kappa}(P(k))U_{\rho(-k'),\kappa'}(-P(k))\frac{1}{\sqrt{4\omega_{\kappa}(P(k))\omega_{\kappa'}(-P(k))}}
\end{aligned}
\end{equation}
\end{widetext}

Finally, we calculate the Matsubara Green's function $\mathcal{G}(k,k',\omega_n)=\int d\omega \frac{A(k,k'\omega)}{i\omega_n-\omega}$, and convolute the $f$-fermion and scalar field Green's function to obtain the $c$-fermion Green's function; note that the $c$-fermion Green's function is gauge invariant, so we can simply set $k=k'$.
\begin{widetext}
\begin{equation}
\begin{aligned}
\mathcal{G}(k,\omega_m)&=\sum_{q,q',\nu_m}\mathcal{G}_f(q,q',\nu_m)\mathcal{G}_f(k+q,k+q',\nu_m)\\
&=\sum_{q,q',\nu_m,\gamma,\kappa}\delta_{P(q),P(q')}\delta_{P(q+k),-P(-q'-k)}V_{\rho(q),\gamma}(P(q))V_{\rho(q'),\gamma}(P(q))\\
&\quad\quad\quad\quad\quad\quad\times U_{\rho(q+k),\gamma}(P(q+k))U_{\rho(-q'-k),\gamma}(-P(q+k))\\
&\quad\quad\quad\quad\quad\quad\times \frac{1}{i\nu_m-\epsilon_\gamma(P(q))}\frac{1}{(\nu_m-\omega_m)^2+\omega_\kappa^2(P(q+k))}\\
&=\sum_{q,q',\nu_m,\gamma,\kappa}\delta_{P(q),P(q')}\delta_{P(q+k),-P(-q'-k)}V_{\rho(q),\gamma}(P(q))V_{\rho(q'),\gamma}(P(q))\\
&\quad\quad\quad\quad\quad\quad\times U_{\rho(q+k),\gamma}(P(q+k))U_{\rho(-q'-k),\gamma}(-P(q+k))\\
&\quad\quad\quad\quad\quad\quad\times \frac{-\beta\omega_\kappa(P(q+k))\tanh(\frac{\beta\epsilon_\gamma(P(q))}{2})+\beta(\epsilon_\gamma(P(q))-i\omega_m)\coth(\frac{\beta\omega_\kappa(P(q+k))}{2})}{2\omega_\kappa(P(q+k))(\omega_\kappa^2(P(q+k))+(\epsilon_\gamma(P(q))-i\omega_m)^2)}
\end{aligned}
\end{equation}
\end{widetext}

The last step is a Matsubara sum which is calculated by the standard way. As an example, the $c$-fermion spectral function from finite size mean-field calculation, with the same temperature and filling compared with that in the QMC simulation inside the Fermi arc phase, is given in Fig.~\ref{fig:fig6}.

\section{Finite Size Analysis of Quasi-particle Weight}
\label{app:FS}
\begin{figure}[htp!]
\includegraphics[width=0.9\columnwidth]{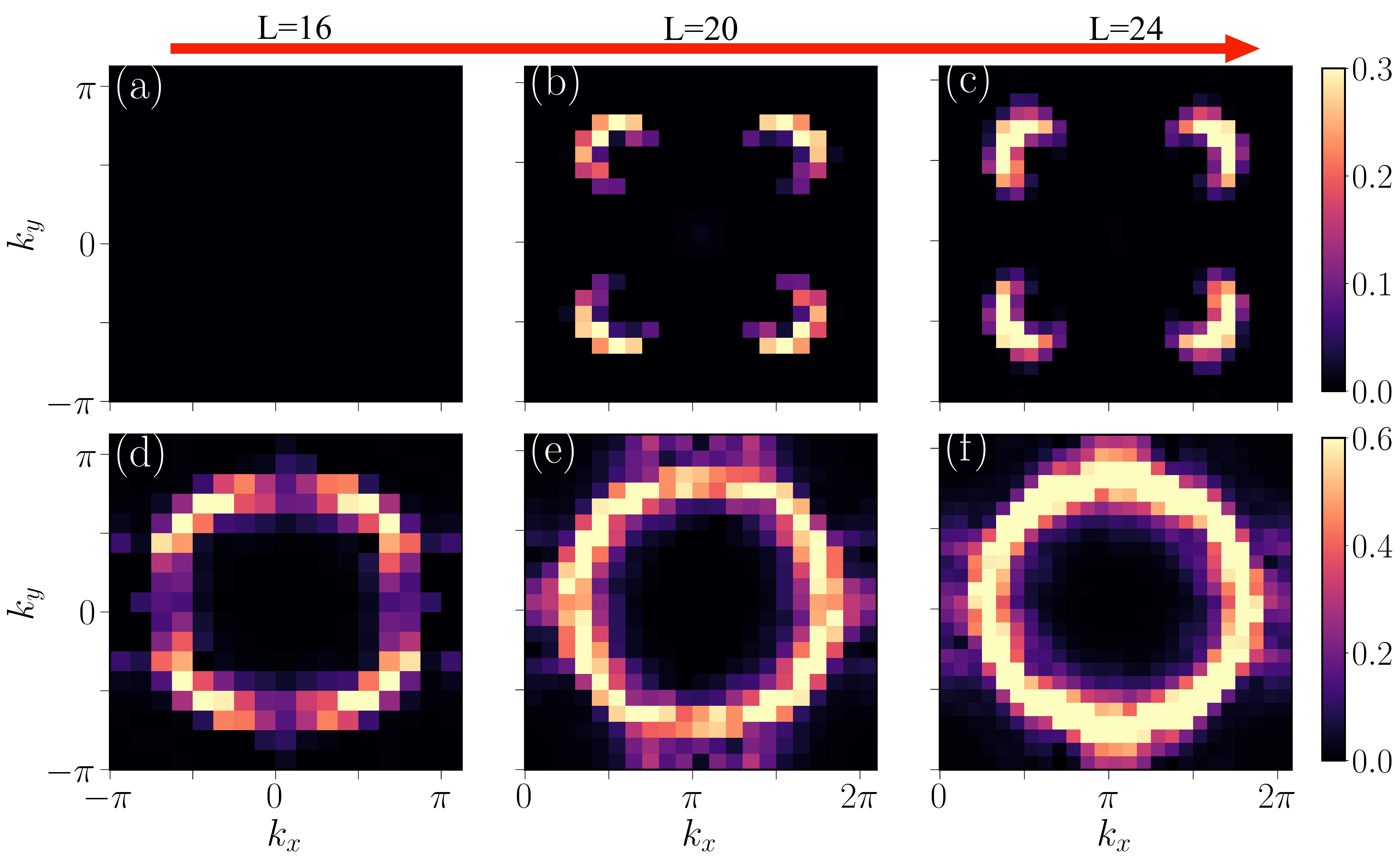}
\caption{Zero-temperature extrapolation of quasiparticle weights $Z_{\mathbf{k}}$. The upper panels are for the Fermi arc phase with parameters $t=0.3, g=0.5, \mu=1.2$, for system sizes $L=16$ (a), $L=20$ (b), and $L=24$ (c). In each panel, the results are shown after extrapolation of different temperatures $\beta=8, 9, 10, 12, 13, 15, 17, 20, 22, 24$. The lower panels are for the deconfined FL phase with parameters $t=1.0, g=0.5, \mu=1.2$, for system sizes $L=16$ (d), $L=20$ (e) and $L=24$ (f) with extrapolation using the largest $\beta=20$.}
\label{fig:fig7}
\end{figure}

In order to get a glimpse of the ultimate fate of our Fermi arc phase from finite-size and finite-temperature simulations to the $L$ and $\beta$ to $\infty$ limit, we perform finite size analysis of the quasi-particle weight $Z_{\mathbf{k}}$. Due to the fact that we only have finite resolution in momentum space and the model is doped away from half filling, the momentum point of interest is, strictly speaking, not a high-symmetry point; therefore we can cannot simply focus on one specific momentum point $\mathbf{k}$ and extrapolate its $Z_{\mathbf{k}}$ to the $L$ to $\infty$ limit. On the other hand, the $T \to 0$ ($\beta \to \infty$) extrapolation is more controllable. Our scheme therefore becomes to perform the zero-temperature extrapolation for each system size first and then to make the comparison between different sizes. The extrapolation itself is done by second-order polynomial regression.

We select the three largest system sizes $L=16, 20, 24$ in DQMC simulations with inverse temperatures $\beta=8, 9, 10, 12, 13, 15, 17, 20, 22, 24$ in the Fermi arc phase ($t=0.3, g=0.5$) and deconfined FL phase ($t=1.0, g=0.5$). The extrapolated results are shown in Fig.~\ref{fig:fig7}. The upper panels are for the Fermi arc phase, and the lower ones are for the deconfined FL. For the cases with the largest lattice size, the antinodal direction in the Fermi arc phase havs vanishing quasi-particle weight in the extrapolated zero temperature Fermi surface. In contrast, the Fermi surfaces of the deconfined FL have persistent quasiparticle weight in the antinodal direction. The important message from the comparison between different sizes is that the quasiparticle weight increases with increasing lattice sizes. In Fig.~\ref{fig:fig7}(a) with $L=16$, $Z_{\mathbf{k}}$ in the BZ all extrapolate to zero, but as the system size increases to $L=20$ and $24$ [Figs.~\ref{fig:fig7}(b) and ~\ref{fig:fig7}(c)], the extrapolated weight along the Fermi arcs becomes more pronounced, verifying the existence of such an exotic metal state in the limit of $L$ and $\beta$ to $\infty$ and that Fermi arcs are not closed pockets when one dopes the Dirac cones of orthogonal fermions. 

\section{Superconductivity}
\label{app:SC}
\begin{figure}[htp!]
\includegraphics[width=0.9\columnwidth]{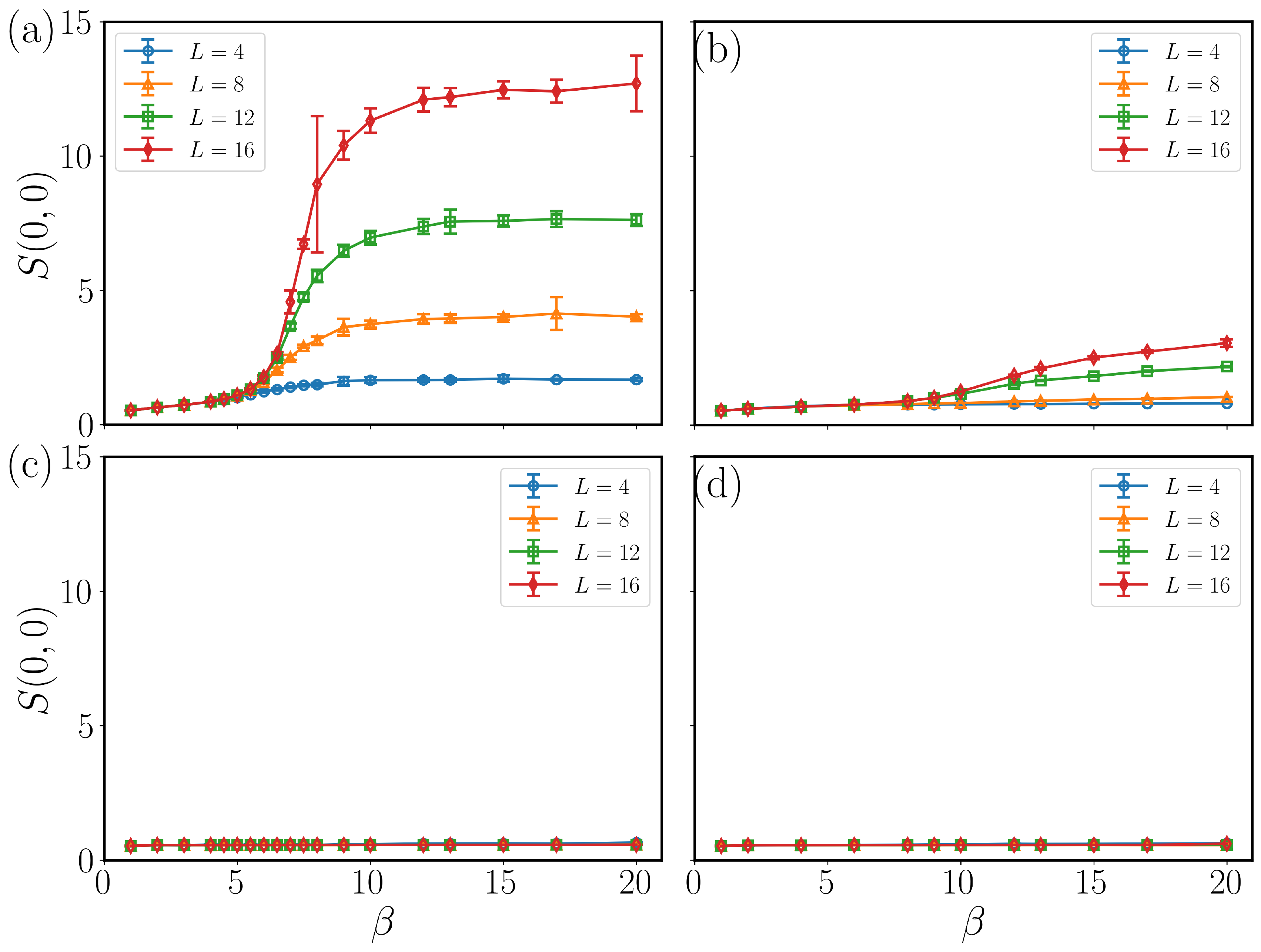}
\caption{ $s$-wave pairing structure factor in (a) $g=1.4, t=0.3$, (b) $g=1.4, t=1.0$, (c) $g=0.5, t=0.3$ and (d) $g=0.5, t=1.0$ corresponding to four circles in the $t-g$ phase diagram in Fig.~\ref{fig:fig2}(a) of the main text. We show the observables for $L=4, 8, 12, 16$ and inverse temperature from $\beta=1$ to $\beta=20$.}
\label{fig:fig8}
\end{figure}

\begin{figure}[htp!]
\includegraphics[width=0.9\columnwidth]{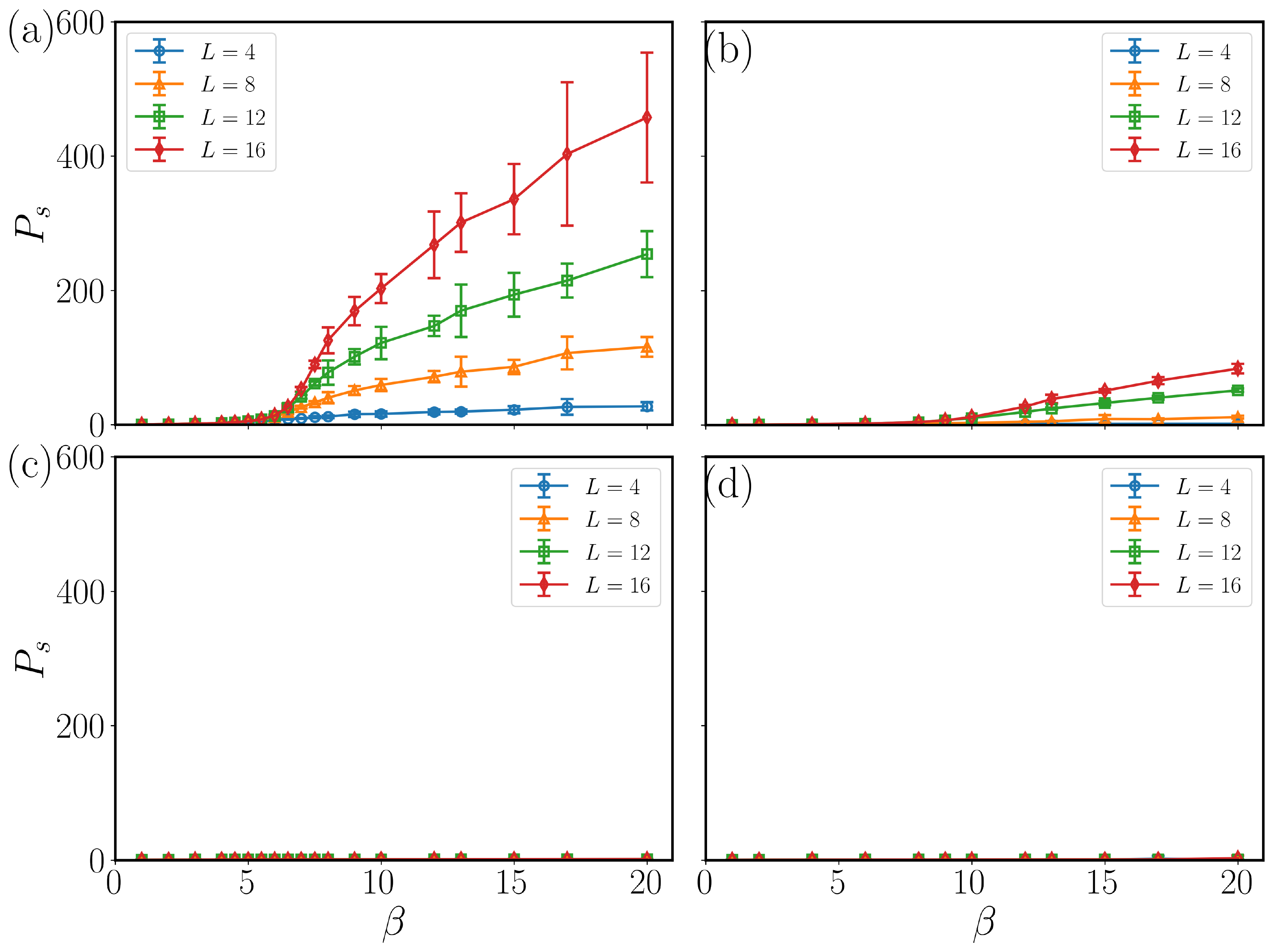}
\caption{ $s$-wave dynamical pairing susceptibility in (a) $g=1.4, t=0.3$, (b) $g=1.4, t=1.0$, (c) $g=0.5, t=0.3$ and (d) $g=0.5, t=1.0$ corresponding to four circles in the $t-g$ phase diagram in Fig.~\ref{fig:fig2}(a) of the main text. We show the observables for $L=4, 8, 12, 16$ and inverse temperature from $\beta=1$ to $\beta=20$.}
\label{fig:fig9}
\end{figure}

We study the $s$-wave superconductivity in our $t-g$ phase diagram. The observables are the $s$-wave pairing structure factor and the dynamical susceptibility defined as follows:
\begin{equation}
	S(\mathbf{k}) = \frac{1}{L^2}\sum_{i,j}e^{-i\mathbf{k} \cdot r_{ij}} \left \langle \Delta^\dagger_i \Delta_j \right \rangle, \; \Delta_i = c_{i\uparrow}c_{i\downarrow}
\end{equation}
and
\begin{equation}
	P_{s}=\frac{1}{L^2} \int_{0}^{\beta} d\tau \left \langle \Delta(\tau)\Delta^{\dagger}(0)+\text{H.c.} \right \rangle, \; \Delta(\tau)=\sum_{i}c_{i\downarrow}(\tau)c_{i\uparrow}(\tau)
\end{equation}

The results are shown in Figs.~\ref{fig:fig8} and \ref{fig:fig9}, respectively. The four panels of each figure are the measurements performed at the solid circle in the phase diagram in Fig.~\ref{fig:fig2}(a) of the main text. In Figs.~\ref{fig:fig8}(a) and ~\ref{fig:fig9}(a), both $S(0,0)$ and $P_s$ show strong enhancement with $L$ and $\beta$, and in the main text, we show the collapse of pairing susceptibility $P_s$ making use of the Kosterlitz-Thouless transition scaling form with $T_c=8.3$. As for Figs.~\ref{fig:fig8}(b) and ~\ref{fig:fig9}(b), this is deep in the confined FL with  $g=1.0, t=1.0$, the temperature at which $S(0,0)$ and $P_s$ start to grow is lower than the $g=1.0, t=0.3$ case, and their absolute values are significantly smaller in comparison with those of $g=1.0, t=0.3$, meaning that although there is superconductivity instability, the $T_c$ is much lower. Attempts to produce the scaling collapse of $P_s$ yield poorer results as $t$ is increased, and from the temperature dependency of the structure factor and pairing susceptibility, we reach the conclusion that the critical temperature of the KT transition versus $t$ is decreasing monotonically. In the Fermi arc and deconfined FL phase, i.e, Figs.~\ref{fig:fig8}(c), ~\ref{fig:fig8}(d), ~\ref{fig:fig9}(c) and ~\ref{fig:fig9}(d), as the temperatures are lowered, we observe no signals of superconductivity in the $s$-wave channel, with similar behaviors in the $d$-wave channel as well, meaning that the deconfined phase in our model does not have superconductivity instabilities.

\section{Explanation of the three phases in the phase diagram}
\label{app:PHASE}

In this appendix, we briefly explain the names of the three phases in the phase diagram in Fig.~2(a) of the main text.
In the main text, the three phases are referred to as the Fermi arc phase, the deconfined Fermi liquid (FL) phase and the confined FL phase, respectively.

The Fermi arc phase and the deconfined FL phase both have a deconfined $\mathbb Z_2$ gauge field and therefore a $\mathbb Z_2$ topological order.
The Ising field $S^z$ is in a disordered state, and the spin-flip excitation is a gapped bosonic quasiparticle carrying a $\mathbb Z_2$ gauge charge.
Therefore, it can be viewed as the $e$ anyon in the $\mathbb Z_2$ topological order.
The $f$ fermion also carries a $\mathbb Z_2$ gauge charge, and it is another fractionalized anyon, which differs from the $e$ anyon by a physical electron $c$.
In other words, the bound state (or the fusion outcome) of an $e$ anyon and a $f$ fermion is the physical electron $c$, which does not carry a gauge charge.
$e$ and $f$ are deconfined anyons in both the Fermi arc and deconfined FL phases, and they differ only in the shape of the Fermi surface of physical electrons:
In the Fermi-arc phase, the $c$ fermion has disconnected Fermi arcs, while in the deconfined FL phase it has a connected large FS.

On the other hand, in the confined FL phase, the $\mathbb Z_2$ gauge field is in the confined phase.
As a result, it has neither topological order nor fractionalized anyon excitations.
Both the $f$ fermion and spin-flip excitation of the Ising field $S^z$ are now confined, and they can only appear together as a $c$ fermion.
Therefore the $c$ fermion is the only low-energy quasiparticle in this phase. Furthermore, it forms a large FS. Therefore this phase is a trivial Fermi liquid with a large FS.

\bibliographystyle{apsrev4-1}
\bibliography{main}

\end{document}